\documentclass[prc,reprint,nofootinbib,preprintnumbers,floatfix,unsortedaddress,superscriptaddress,longbibliography,times]{revtex4-2}

\usepackage{hyperref}
\hypersetup{unicode, pdfprintscaling=None, colorlinks}

\usepackage{orcidlink}
\usepackage{multirow}
\usepackage{siunitx}
\usepackage{colortbl}
\usepackage{dcolumn}% Align table columns on decimal point

\usepackage{tikz}
\usetikzlibrary{decorations.markings, arrows, decorations.pathmorphing}

\usepackage{graphicx}
\usepackage{latexsym}
\usepackage{amsmath,amssymb,amsthm}
\usepackage{amsfonts}
\usepackage{bm}
\usepackage{bbm}
\usepackage{microtype}
\usepackage{xspace}
\usepackage{placeins}

\usepackage{hyperref}
\hypersetup{
    colorlinks=true,
    linkcolor=blue,  
    citecolor=blue,
    urlcolor=magenta,
    }
\usepackage[capitalise]{cleveref}

\usepackage{mathtools}

\usepackage[export]{adjustbox}

\newlength{\FigureWidth}
\usepackage{xcolor}

\usepackage{makecell}
\usepackage{tabularx}
\newcolumntype{L}[1]{>{\raggedright\arraybackslash}p{#1}}
\newcolumntype{C}[1]{>{\centering\arraybackslash}p{#1}}
\newcolumntype{R}[1]{>{\raggedleft\arraybackslash}p{#1}}
\newcommand\TopRule{\Xhline{0.08em}}

\newcommand\MidRule{\Xhline{0.03em}}
\newcommand\BotRule{\Xhline{0.08em}}

\newtheorem{Th}{Theorem}[section]
\newtheorem{Lemma}[Th]{Lemma}

\newcommand{\HNN}{{H_{NNGMM}}}
\newcommand{\HMG}{{H_{MGGMM}}}

\newcommand{\unitmat}{{\mathbbm{1}}}

\def\a{\alpha}

\def\d{\delta}
\def\e{\epsilon}

\def\g{\gamma}

\def\j{\psi}
\def\k{\kappa}
\def\l{\lambda}
\def\m{\mu}
\def\n{\nu}

\def\p{\pi}
\def\q{\theta}

\def\s{\sigma}

\def\D{\Delta}

\def\L{\Lambda}

\def\co{{\cal O}}

\def\car{{\cal R}}

\def\O{{\mathrm{O}}}
\def\SO{{\mathrm{SO}}}
\def\Spin{{\mathrm{Spin}}}
\def\Pin{{\mathrm{Pin}}}
\def\SU{{\mathrm{SU}}}

\def\bo{{\raise.15ex\hbox{\large$\Box$}}}               % D'Alembertian
\def\TH{{\raise.2ex\hbox{$\displaystyle \bigodot$}\mskip-4.7mu \llap H \;}}
\def\face{{\raise.2ex\hbox{$\displaystyle \bigodot$}\mskip-2.2mu \llap {$\ddot\smile$}}} % happy face
\def\dg{\sp\dagger}% hermitian conjugate
\def\sp#1{{}^{#1}}                              % superscript (unaligned)
\def\ket#1{\left| #1\right\rangle}              % | >
\def\leftrightarrowfill{$\mathsurround=0pt \mathord\leftarrow \mkern-6mu
        \cleaders\hbox{$\mkern-2mu \mathord- \mkern-2mu$}\hfill
        \mkern-6mu \mathord\rightarrow$}
\def\dvec#1{\vbox{\ialign{##\crcr
        \leftrightarrowfill\crcr\noalign{\kern-1pt\nointerlineskip}
        $\hfil\displaystyle{#1}\hfil$\crcr}}}           % <--> accent
\def\ha{\frac12}                                        % 1/2
\def\sfrac#1#2{{\vphantom1\smash{\lower.5ex\hbox{\small$#1$}}\over
        \vphantom1\smash{\raise.4ex\hbox{\small$#2$}}}} % alternate fraction
\def\bfrac#1#2{{\vphantom1\smash{\lower.5ex\hbox{$#1$}}\over
        \vphantom1\smash{\raise.3ex\hbox{$#2$}}}}       % "
\def\afrac#1#2{{\vphantom1\smash{\lower.5ex\hbox{$#1$}}\over#2}}    % "
\begin{document}

\title{Transverse Field \texorpdfstring{$\gamma$}{Lg}-Matrix Spin Chains}
\author{Rui Xian Siew\orcidlink{0000-0002-0745-8853}}
\email{ruixian.siew@duke.edu}
\affiliation{ Department of Physics, Box 90305, Duke University, Durham, North Carolina 27708, USA}
\author{Shailesh Chandrasekharan\orcidlink{0000-0002-3711-4998}}
\email{sch27@duke.edu}
\affiliation{ Department of Physics, Box 90305, Duke University, Durham, North Carolina 27708, USA}
\author{Ribhu K. Kaul\orcidlink{0000-0003-1301-7744}}
\email{ribhu.kaul@psu.edu}
\affiliation{Department of Physics, The Pennsylvania State University, University Park, PA 16802, USA}

\begin{abstract}
We introduce a simple lattice spin model that is written in terms of the well-known four-dimensional $\gamma$-matrix representation of the Clifford algebra. The local spins with a four-dimensional Hilbert space transform in a spinorial  $(1/2,0) \oplus (0,1/2)$ representation of SO(4), a symmetry of our model. When studied on a chain, and as a function of a transverse field tuning parameter, our model undergoes a quantum phase transition from a valence bond solid phase to a critical phase that is described by an $\SU(2)_1$ WZW field theory.  \end{abstract}

\maketitle

\section{Introduction}
\label{sec-1-alt}

Understanding the non-perturbative properties of quantum field theories is challenging in theoretical physics. These challenges become even more difficult when the path integrals depend on the topological classification of the field configurations. Often the topological information encodes the anomalous symmetries of the theory and appears through topological terms in the classical action of the quantum field theory \cite{Wess:1971yu,Callan:1976je,Witten:1983tw}.
Understanding the physics of such topological terms has become an exciting area of research at the crossroads of both high energy and condensed matter physics \cite{RevModPhys.88.035001,altlandsimons}. In the context of high energy physics, the value of the $\theta$-parameter in QCD is found to be unnaturally small,  which is a puzzle referred to as the strong-CP problem \cite{Peccei:1977hh}. In condensed matter physics, $\theta$-terms play a central role in our understanding of novel phenomena related to spin liquids, deconfined criticality, symmetry-protected topological phases, and localization, among others (see \cite{Senthil:2005jk,VS:2013,slagle:2015}).

There are very few non-perturbative methods to study the physics of quantum field theories with topological terms since they often lead to complex actions even when path integrals are written in Euclidean space. This makes it difficult to use Monte Carlo methods unless special tricks are designed to solve the associated sign problems \cite{Bietenholz:1995zk}. On the other hand, if we take inspiration from condensed matter physics, there are examples of quantum lattice models that are free of sign problems when formulated properly, while also describing quantum field theories with emergent topological terms at long distances. A well-known example of this is the one-dimensional quantum spin-half chain. In this case, there is extensive literature that shows that the model is described by the $k=1$ $\SU(2)$ Wess-Zumino-Witten (WZW) model with a marginally irrelevant coupling at long distances \cite{Affleck:1987ch}. Recently it has also been demonstrated that spin-ladders naturally reproduce the physics of the $\O(3)$ non-linear sigma model with a $\theta$-term in the continuum, all the way from the UV fixed point to the IR fixed point \cite{Caspar:2022llo}.

The idea of being able to construct any desired continuum quantum field theory via the critical physics of a quantum lattice Hamiltonian with a proper choice of a finite-dimensional local Hilbert space, is becoming a new area of research and is referred to as qubit regularization \cite{Singh:2019uwd,Bhattacharya:2020gpm,Zhou:2021qpm,Liu:2021tef}. If these new quantum lattice models provide an alternate approach to studying continuum quantum field theories, they can then be studied using quantum computers when they become available \cite{Banuls:2019bmf}. Since anomalies arise through projective representations in the quantum Hilbert space \cite{Nelson:1984gu}, it is possible that when the quantum Hilbert space of the lattice model is realized through these representations, topological terms may naturally be induced in the Euclidean effective actions of the long-distance theory that arise at some quantum critical points. For example, when the $\SO(3)$ symmetry is realized through a quantum Hilbert space that contains spinorial representations, even simple quantum lattice models naturally contain Wess-Zumino-Witten (WZW) terms \cite{Bhattacharya:2023wuz}.

The motivation of our current work is to explore new quantum lattice models with $\SO(4)$ symmetries realized through Hilbert spaces using spinorial representations. We hope that these different representations can help us identify exotic quantum critical points where quantum field theories with new types of topological terms will naturally emerge. Traditional lattice models with $\SO(4)$ symmetries are constructed using local Hilbert spaces in the vectorial representations. These describe a quantum particle moving on $S^3$ (unit sphere in four-dimensional Euclidean space) and thus contain the sum of irreducible representations (irreps) of $\mathfrak{so}(4) \equiv \mathfrak{su}(2)_+\oplus \mathfrak{su}(2)_-$ of the form $(s_+,s_-)=(s,s)$, where $s=0,1/2,....$ are the usual irreps of $\SU(2)$. 
However, we can formulate lattice models using irreps of the universal covering group of $\SO(4)$ which is $\Spin(4)$. Then $\SO(4)$ is realized projectively through a two-to-one mapping. The simplest non-trivial $\Spin(4)$ representation is of the form $(1/2,0) \oplus (0,1/2)$. Thus, it is natural to explore $\SO(4)$ symmetric quantum lattice models realized via a Hilbert space that contains this representation.  
A natural question is whether such lattice models contain exotic quantum critical points with novel topological terms.
Since these models have a spinorial representation, we can extend the Lieb-Shultz-Mattis (LSM) theorem to these spin models, which forbid simple gapped symmetry-preserving phases. This, in turn, implies that the effective field theory cannot be described by simple sigma models with disordered phases, and will likely contain topological terms. One such term we are interested in is the $\theta$-term of the $\SO(4)$ model in $2+1$ dimensions. We note that we may need a quantum lattice Hamiltonian where the $\SO(4)$ symmetry of the $\theta$-term only arises as an emergent symmetry, since anomalies may prevent the realization of the SO(4) symmetry as an on-site symmetry \cite{SenthilPC,wangso5,Nguyen:2022aaq}.

With this general motivation, our specific goal in this work is to introduce and study new many-body quantum spin Hamiltonians invariant under a $\SO(4)$ global symmetry, realized using a local four-dimensional Hilbert space on every site that transforms under the reducible representation $(1/2,0) \oplus (0,1/2)$ of the symmetry group. In a way, our lattice models are extensions of quantum spin-half models that are invariant under $\Spin(3)\simeq \SU(2)$ realized using spinorial representations to $\Spin(4)$. As we explain here, this can naturally be achieved by replacing the three Pauli matrices with the five Dirac (gamma) matrices as the basic quantum operators on each lattice site. Hence, we call our models $\g$-matrix models (or GMM). %While our motivation to construct these new types of models is mainly theoretical, we note that models with high spin symmetry can be realized approximately using cold atomic systems \cite{PhysRevLett.91.186402,ConjunWu2006}. 

 In this work, we focus on a simple nearest-neighbor Hamiltonian in one spatial dimension. This allows us to understand how to study the interesting quantum many-body physics of these new spin models, using both analytic arguments and controlled numerical methods such as exact diagonalization and the density matrix renormalization group (DMRG) on large system sizes. In future studies, we plan to study the GMMs in higher dimensions.
We find that the simple one-dimensional model studied in this work hosts the quantum phase transition between a dimer phase and a critical phase, which is in the same universality class as the one in the $J_1-J_2$ model of a spin-half anti-ferromagnetic chain with a next-to-nearest neighbor coupling \cite{Okamoto1992,affleck1988}. It is interesting that the frustration arising from a next-to-nearest neighbor coupling in the $J_1-J_2$ model is naturally induced through a simple nearest neighbor coupling in the GMM model.
While establishing this phase diagram, we generalize two well-known results about the $S=1/2$ chain~\cite{auerbach} to the $\gamma$-matrix spin model: (1) our GMM satisfies the LSM theorem which proves that it cannot have a trivial gapped phase, and (2) in an extension of the famous Majumdar-Ghosh point, for a special ratio of the nearest and next-nearest neighbor GMM interaction, the model is exactly solved with a dimerized ground state.
%\url{https://wucj.physics.ucsd.edu/research/highsymmetry/sp4.html}. People have studied fermionic systems with SO(4) and SO(5) symmetries. Half-quantum vortices carry SO(4) quantum numbers.

%\SRX{ a reminder to emphasize the model is different from others in that our local Hilbert space is a direct sum of symmetry group irreps instead of tensor product.}

Our paper is organized as follows. In \cref{sec-2}, we introduce the model we study in our work. In \cref{sec-3}, we provide strong evidence that our model has two phases: a dimerized phase and a critical phase separated by a quantum phase transition, all of the same type as  \cite{Okamoto1992}. In \cref{sec-4}, we show results using exact diagonalization and tensor network calculations that confirm our predictions. In \cref{sec-5}, we present our conclusions. Finally, we have included four appendices, which are dedicated to reviews of well-known facts and details that are too technical for the main manuscript.
\section{The Model}
\label{sec-2}

{\em Hilbert Space:} Here we will consider a quantum spin model in which the local on-site Hilbert space is four-dimensional, i.e. there are four states on each site of the lattice\footnote{In contrast, the most popular spin models (Heisenberg, transverse field Ising) have a two-dimensional local Hilbert space.}. The Hilbert space of the many-body lattice system is built up in the usual way by tensor products so that with $L$ sites, the full Hilbert space has a dimension $4^L$. We note that spin models with four-dimensional on-site Hilbert spaces have been studied in the past in the context of the $S=3/2$ representation of SU(2), as well as fundamental representations of SU(4) and SO(4). As we shall see, the model we introduce here is different from these. In \cref{app-1}, we review how the four-dimensional Hilbert space can be viewed as a reducible spinorial representation of the $\Spin(4)$ symmetry.
%and \RKK{drop SO(5) and focus on SO(4)?} $\SO(5)$ symmetries, which we refer to as $\Spin(4)$ and $\Spin(5)$ respectively. These are naturally realized with a four-dimensional Hilbert space 

{\em Operators:} 
%Before we introduce our specific model, we discuss the most general four-dimensional on-site Hilbert space models. 
Spin models are constructed from local operators that act on the site Hilbert space. From the perspective of the $\Spin(4)$ symmetry, it is natural to work with the five Hermitian $4\times 4$ anti-commuting Dirac matrices (in contrast to the three $2\times 2$ Hermitian anti-commuting Pauli matrices that are used for the usual two-dimensional on-site Hilbert space models). We label these matrices as $\g^\m ~(\m=1,2,3,4,5)$ where $\g^5=-\g^1\g^2\g^3\g^4$ and they satisfy the anti-commutation relations
\begin{align}
    \{\g^\m,\g^\n\}=2\d^{\m\n}.
\end{align} 
As we explain in \cref{app-1}, the five Dirac matrices transform as a 5-vector under the spinorial representations of $\Spin(5)$. Thus, the Dirac matrices are similar to the Pauli matrices which transform as a 3-vector under the spinorial representation of $\SO(3)$ or $\Spin(3)$ transformations. However, unlike the three Pauli matrices and the identity matrix that span the space of $2\times 2$ Hermitian matrices, the five $\gamma$-matrices plus the identity matrix do not form a complete basis for $4\times 4$ Hermitian matrices; we have to include in addition the 10 commutators of $\gamma$-matrices, $\sigma_{\mu\nu}$.  
This means we can construct several interesting spin Hamiltonians using the Dirac matrices with SO(4) symmetry realized in the spinorial representation, which we collectively refer to here as {\em $\g$-matrix models}.\footnote{We note that any spin model with a local four dimensional Hilbert space can be written in terms of $\gamma$-matrices and their commutators. Some past work has addressed such models, see e.g. \cite{PhysRevB.108.094411,PhysRevLett.102.217202,PhysRevB.87.041103,Chulliparambil2020:16fway}. Here we focus exclusively on the SO(4) symmetry and its spinorial representation.} Some of these models and their symmetry properties are discussed in \cref{app-2}. %\SRX{Consider removing the previous sentence if we do not plan to show other GMM. Also, for consistency, we should decide whether/when to introduce the acronym GMM.}
These models can be defined in any dimension and we expect that they host rich phase diagrams that have not yet been explored.  In this work, we focus on one of the simplest and perhaps most natural of these models in one spatial dimension, which we now introduce.  %and show that this simple model already contains an interesting quantum phase transition between a critical phase whose long-distance physics is described by the $k=1$ WZW CFT and a massive dimerized phase obtained due to spontaneous breaking of translation symmetry.

{\em Transverse field $\gamma$-matrix model:}  A simple symmetric interaction we can introduce between two spins at sites $i$ and $j$ is $\sum^4_{\mu=1}\gamma^{\mu}_i \gamma^{\mu}_j$.  However, in this case, we still have an extra $\gamma^5$ term, which has not appeared in the exchange interaction and to which we can couple an external ``transverse" field. Putting all this together, we write down the following one-dimensional spin chain model,
\begin{align}
    H=\sum_{j=1}^{L}\left[J\ \sum^4_{\mu=1}\g^\m_j\g^\m_{j+1}\ +\ h\ \g^5_j\right]\,.
\label{eq:TFGMM-1}
\end{align}
which we refer to as the transverse field O(4) $\g$-matrix model (TFGMM). While this model can be written down in any dimension, in this work we analyze it extensively on a one-dimensional chain, where we are able to obtain its full phase diagram. We will use both periodic boundary conditions (PBC) and open boundary conditions (OBC) to solve the system. In the PBC case, the term $\g^\m_L\g^\m_{L+1}$ in \cref{eq:TFGMM-1} is replaced by $\g^\m_L\g^\m_1$, whereas the term is dropped in the OBC case.

It is useful to note that the sign of $h$ and $J$ can be changed by performing $\mathbb{Z}_2$ unitary transformation on $H$ using
\begin{align}
    \car_{51}=\prod_{j}\ (i\g^5_j\g^1_j)
\label{eq:z1}
\end{align}
and
\begin{align}
    \car_5^{odd}=\prod_{j \in\text{odd}}\g^{5}_j\,,
\label{eq:z2}
\end{align}
implying that these signs do not change the physics. For this reason, we will assume $J, h \geq 0$ in this work. Further, if we measure energies in units of $\sqrt{J^2+h^2}$, it is natural to define the dimensionless Hamiltonian 
\begin{align}
    \hat{H}\ =\ \sum_{j=1}^{L}\left[\sqrt{1-\alpha^2}\ \sum_{\m = 1}^4 \g^\m_j\g^\m_{j+1}\ +\ \alpha\ \g^5_j\right]\,.
\label{eq:TFGMM-2}
\end{align}
where we have defined a dimensionless coupling $\a=h/\sqrt{J^2+h^2}, 0\le \a \le 1$ to probe the effects of the two terms. Further, since $\{H,\car_{51}\car^{odd}_{5}\}=0$, the spectrum of $H$ is symmetric about zero.

 We now analyze the symmetries of the TFGMM. The nearest-neighbor term in \cref{eq:TFGMM-1} with coefficient $J$ is invariant under the $\O(4)$ symmetry. The transverse field term with coefficient $h$ breaks the $\O(4)$ symmetry to $\SO(4)$ and splits the two $\SU(2)$ sectors in the $\SO(4)$ symmetry group (see \cref{app-1} for more details). Hence, at very large $h$ (or equivalently when $\alpha=1$ in \cref{eq:TFGMM-2}) our model favors one $\SU(2)$ sector, and, as we show below, the model can be described by the physics of the $\SU(2)$ Heisenberg chain.

%\RKK{edit!} The global symmetry group is $\mathrm{Spin}(4)$.
%For example, the involutive unitary operators below are elements of $\mathrm{Spin}(4)$:
%\begin{align}
%    \car_{5}=\prod_j \g^5_j\,,~~\car_{\m\n}=\prod_j 2\s^{\m\n}_j\,.
%\end{align}
%The occupation numbers of spin $\s=\uparrow,\downarrow$ at a site $j$ is defined to be 
%\begin{align}
    %\hat{n}_{j\uparrow}=\ha+\s^{12}_j\,,~~\hat{n}_{j\downarrow}=\ha+\s^{34}_j\,.\label{eq: n as sigma}
%\end{align}
%Then, the conservation of total occupation numbers
%\begin{align}
 %   \hat{n}_{\a}=\sum_{j}\hat{n}_{j\s}
%\end{align}
%is a consequence of the global $\mathrm{Spin}(4)$ symmetry.
%\footnote{The definitions \ref{eq: n as sigma} are not unmotivated. One can define fermions $\hat{c}_{j\s}$ from linear combinations of $\g^\m_j$ and a non-local chain of $\g^5$, and then define $\hat{n}_{j\s}=\hat{c}\dg_{j\s}\hat{c}_{j\s}$.}

%For sufficiently large finite $L$ and sufficiently small $\a>0$, the lowest two energy eigenstates are $\mathrm{Spin}(4)$ singlets. In the thermodynamic limit, these two singlets correspond to the doubly-degenerate ground state of $\hat{H}_4$ that spontaneously breaks translation symmetry $T$.
%Thus, the system is in the VBS state for small $\a$ in that limit.
%At the other extreme of $\a\approx 1$, the low energy theory is approximately that of the $J_1-J_2$ model, which has a spin triplet instead of a singlet as the first excited state.

{\em 2-site problem:} Before tackling the many-body problem on the chain, it is helpful to gain some intuition by studying the spectrum of the 2-site model, i.e. \cref{eq:TFGMM-1} with $L=2$ under OBC. Explicitly, the Hamiltonian is
\begin{align}
    H=\sqrt{1-\a^2}\sum_{\m=1}^{4}\g^\m_1\g^\m_2+\a(\g^5_1+\g^5_2)\,.
\end{align}
We will try to understand the irreducible representations of its eigenstates under $\Spin(4)$ symmetry, as well as their energetics. As we explain in \cref{app-1}, irreducible representations of $\mathrm{Spin}(4)=\SU(2)_+\times\SU(2)_-$, can be understood as a tensor product of two $\mathrm{SU}(2)$ irreps, which can be labeled by $(s_+,s_-)$, where $s_{\pm}\in\ha \mathbb{Z}$. Since our model is invariant under $\Spin(4)$, all the eigenstates of the two-site Hamiltonian can be classified through the quantum numbers $(s_+,m_+;s_-,m_-)$, where $s_{\pm},m_{\pm}\in \ha \mathbb{Z}$ such that $-s_{\pm}\le m_{\pm}\le s_{\pm}$. The 16 energy eigenstates split into two non-degenerate singlets $(0,0)$ with eigenvalues 
$E=\pm 2\sqrt{4-3\a^2}$, one $\mathrm{SU}(2)_+$ triplet with eigenvalue $2\alpha$, one $\mathrm{SU}(2)_-$ triplet with eigenvalue $-2\alpha$, and two 4-vectors (with $s_+=s_-=\ha$) with eigenvalues $\pm 2\sqrt{1-\a^2}$. 

Since $\{H,\car_{51}\car_5^{odd}\}=0$, the 2-site spectrum is invariant under $\hat{H}\rightarrow-\hat{H}$, i.e., for every positive energy eigenstate, there is also a negative energy eigenstate. Also, as we explain in \cref{app-1}, when $\alpha=0$ there is a $\mathbb{Z}_2$ symmetry that flips between the two spins $\SU(2)_+ \leftrightarrow \SU(2)_-$. 
The ground state is an eigenstate of this additional symmetry.
Notably, the ground state is always a singlet under $\Spin(4)$ for all $\a$. This feature resembles the usual antiferromagnetic model where the exchange interaction favors the formation of two-site singlets.
We note that at $\a=1$ the singlet becomes degenerate with a triplet, giving four ground states. This is exactly what we expect since the transverse field term by itself gives two degenerate ground states per site.

%and $0\le s_++s_-\le \frac{L}{2}$. 

% We will label the 2-site problem's energy eigenstates in the following form: $\ket{E;s_+,m_+;s_-,m_-}$, where $E$ is the energy and the other data label the Spin(4) representation. The 16 states of the 2-site system have the following eigenstates,
% \begin{align}
%    \text{2 singlets:}~~ \ket{\pm 2\sqrt{4-3\a^2};0,0;0,0}\,,\\
%    \text{2 triplets:}~~ \ket{ 2\a;1,m_+;0,0}\,,~~\ket{ -2\a;0,0;1,m_-}\,,\\
%    \text{2 4-vectors:}~~ \ket{ \pm 4\sqrt{1-\a^2};\ha,m_+;\ha,m_-}\,.
% \end{align}
\begin{figure}[t]
\centering    \includegraphics[width=0.5\textwidth]{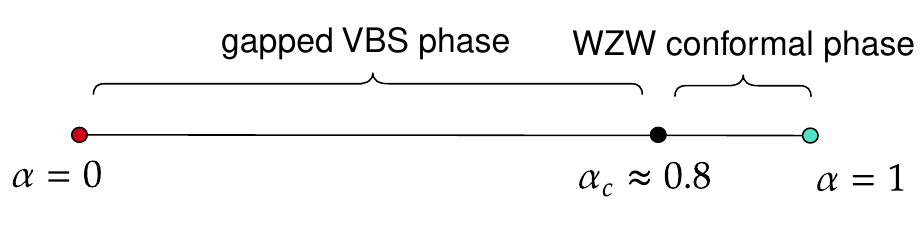}
\caption{ \label{fig: phase diagram} Phase diagram of the TFGMM, given in \cref{eq:TFGMM-1}. The critical point $\alpha_c$ separates a VBS phase from the $k=1$ WZW phase. Since the point $\alpha=1$ consists of decoupled sites, it is extensively degenerate and unphysical. In Sec.~\ref{sec-3} we present arguments to infer this phase diagram and in Sec.~\ref{sec-4} we present numerical simulations that corroborate these inferences.}
\end{figure}   

\section{Phase Diagram}
\label{sec-3}
This section puts forth general arguments using exactly solvable models and perturbation theory to establish the phase diagram of \cref{eq:TFGMM-1} (See \cref{fig: phase diagram}). The upshot of our reasoning is that the model is valence-bond-solid (VBS) ordered (breaking translational symmetry) for small values of $\alpha$, and it enters the $k=1$ WZW critical phase for $\alpha > \alpha_c$ where $\alpha_c < 1$ is a quantum critical point. The universality class of the long-distance physics near $\alpha_c$ is the same as near the quantum critical point in $J_1$-$J_2$ Heisenberg chain, where a coupling switches from being marginally irrelevant to marginally relevant. In the next section, i.e. \cref{sec-4}, we provide detailed DMRG and exact diagonalization results that substantiate this picture. 

\subsection{\texorpdfstring{$\a \lesssim 1$}{Lg}: Degenerate Perturbation Theory}\label{subsec: degen pert}
At $\a=1$, the TFGMM has only the transverse field term and it is trivially solvable. Since $\gamma_i^5$ has two eigenvalues which are each two-fold degenerate on each site, the many-body ground state is highly degenerate and spans a $2^L$-dimensional subspace whose energy is $E_0=-L$. The two degenerate states at each site transform under the irrep $(s_+=0,s_-=\ha)$ of $\Spin(4)$, which implies that dynamically the $\Spin(4)$ symmetry has reduced to the  $\SU(2)_-$ symmetry. When we deviate from this strict $\a=1$ limit, we expect the degeneracy to be lifted and a unique quantum ground state to be selected on finite lattices. In this regime (i.e., when $\a \lesssim 1$), we can use degenerate perturbation theory to argue that our model maps to the famous spin-$\ha$ $J_1$-$J_2$ Heisenberg chain. The effective Hamiltonian up to quartic order of the small parameter $\e=\sqrt{1-\a^2}$ is given by
\begin{align}
     \hat{H}_{\text{eff}}
     \ =\  & L\left(-1-\ha \e^2+\frac{7}{8}\e^4\right) \nonumber \\
     & +\ \e^2 \left(1-\frac{3}{2}\e^2\right)\sum_{j}(\s^i)_j(\s^i)_{j+1}\nonumber\\
     &\ +\frac{3\e^4}{4}\sum_{j}(\s^i)_{j}(\s^i)_{j+2}+\co(\e^5)\,.
\end{align}
The leading order term in this effective Hamiltonian is simply the nearest neighbor Heisenberg exchange coupling $J_1$ and the next-to-leading order term is 
the next-nearest neighbor coupling $J_2$. We can identify
\begin{align}
J_1 =  4 \e^2 \left(1-\frac{3}{2}\e^2\right),\quad J_2 =  3\e^4. 
\end{align}
This allows us to conclude quite reliably that when close enough to $\alpha=1$, the system must be in the same phase as the nearest neighbor Heisenberg spin chain. As we move away from this limit, the ratio of $J_2/J_1$ increases. Based on our knowledge of the Heisenberg chain, we expect a transition to a VBS phase when $J_2/J_1\approx 0.2411$ \cite{Okamoto1992}. Undoubtedly, our perturbative expansion is not controlled and may not be valid in this regime. Nevertheless, we can still use our leading order expressions above to estimate the value of $\alpha_c$ where the transition from the WZW critical phase to the VBS occurs. This gives us $\a_c \approx 0.89$. We will find in Sec.~\ref{sec-4} from large-scale numerics that this estimate is in rough agreement with the numerically determined transition point, validating the picture emerging from perturbation theory. %$\e\approx \pm 0.46. %The Majumdar-Ghosh model corresponds to $\a\approx \pm 0.82$ or $\e\approx \pm 0.58$. 

\subsection{\texorpdfstring{$\a \ll 1$}{Lg}: Majumdar-Ghosh type analysis}
\label{subsec: gapped phase}

We now consider the limit of $\a=0$ by switching off the transverse field. While we will need to resort to large-scale numerical methods to ultimately determine the ground state at this point, we can gain insight by deforming our model to obtain an exactly solvable point akin to the famous Majumdar-Ghosh (MG) point of $S=1/2$. Inspired by their work, we deform our model by adding a second neighbor interaction to obtain the $J$-$J'$ model
\begin{align}
\label{eq:jjprime}
\hat{H}\ =\ \sum_{j=1}^{L}\sum_{\m=1}^4\ \left(J \g^\m_j\g^\m_{j+1}+J^\prime\g^\m_j\g^\m_{j+2}\right)\,,~~J,J'\ge 0\,.
\end{align}
In \cref{app-3}, we show that when $J^\prime/J=1/2$, the ground state of this deformed model is exactly solvable and it is dimerized with two-fold degeneracy. This means at least when $J^\prime/J=1/2$ the model is indeed in the VBS phase.

\begin{figure}[t]
\centering
\includegraphics[width=0.5\textwidth]{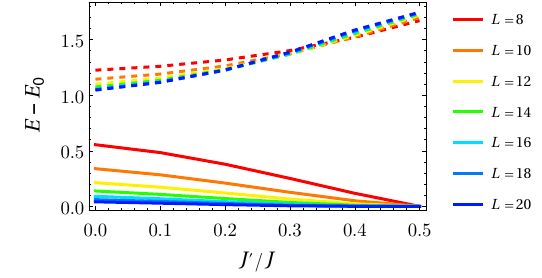}
\caption{
% \RKK{should we maybe show fewer L's? Also maybe good to include the $J^\prime=0$ point? Also maybe label the x-axis $J^\prime/J$? Ideally the font used for the labels on the plot should be the same as the latex font. This can be set in Mathematica. also good to generally follow consistent color schemes throughtout the paper, if possible.}
Lowest three energy eigenvalues for the $J$-$J^\prime$ model of Eq.~\ref{eq:jjprime}. We fix $J=1$, and increase $J^\prime$ from $0.0$ to $0.5$. Solid lines are for $E_1-E_0$, whereas dashed lines are for $E_2-E_0$. Dashed and solid lines of the same color have the same $L$. This numerical analysis allows us to conclude that the exactly solvable MG point $J^\prime/J=0.5$ and Eq.~\ref{eq:TFGMM-1} at $\alpha=0$ are in the same phase, i.e. they are both in a valence-bond-solid ordered phase. }
\label{fig: NNN GMM gap}
\end{figure}
 
%At the MG point, the $J$-$J'$ model is a sum of the 2-bond operator 
%\begin{align}
%K_j\ =\ \sum_{\m=1}^4\ (\g^\m_j+\g^\m_{j-1}+\g^\m_{j+1})^2   -4
%\end{align} 
%over all site index $j$, up to an additive constant. This positive Hermitian operator $K_j$ has zero-eigenstates. Thus, any ground state $\ket{\j_0}$ of the $J$-$J'$ model at the MG point must satisfy $K_j\ket{\j_0}=0$ for any $j$. It can be shown that this is only possible if $\ket{\j_0}$ is a tensor product of Spin(4) singlets $\ket{i;j}$ such that $|i-j|\le 2$, where $\ket{i;j}$ is the unique ground state of the 2-site GMM $\g^\m_i \g^\m_j$. For $L=4$, there are 3 ways to form such tensor products: $\ket{1;2}\otimes \ket{3;4}$, $\ket{2;3}\otimes \ket{4;1}$, and $\ket{1;3}\otimes \ket{2;4}$, so the $L=4$ theory has a trimerized ground state. For even $L>4$, there are only two possibilities: $\bigotimes_{k=1}^{L/2}\ket{2k;2k+1}$ and $\bigotimes_{k=1}^{L/2}\ket{2k-1;2k}$. We conclude that the $J$-$J'$ model at MG point for even $L>4$ has an exactly dimerized 2-fold degenerate ground state.

\begin{figure*}[t]
\centering
\includegraphics[width=0.45\textwidth]{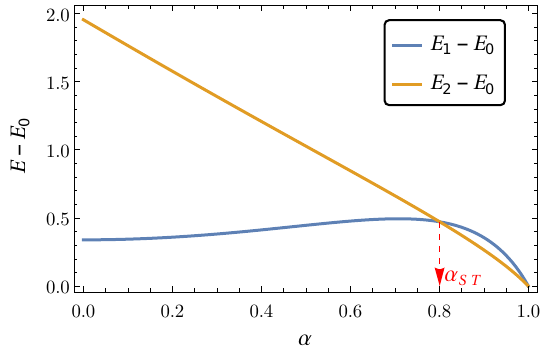}
\includegraphics[width=0.45\textwidth]{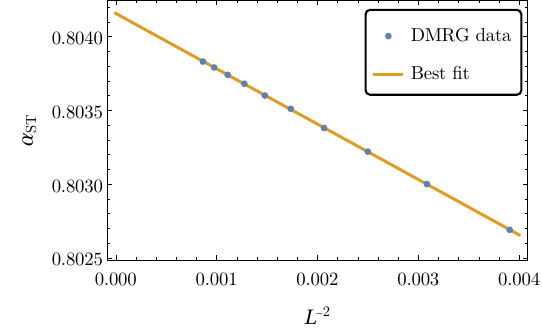}
\caption{In the left figure we plot the energy gap between the lowest singlet state and the ground state and lowest triplet state and the ground state, obtained using exact diagonalization of TFGMM for $L=10$, as a function of $\a$. At the transition between the VBS phase and $k=1$ WZW phase, the two gaps are supposed to be equal. Here we find that this occurs at around $\a_{ST}\approx 0.801$. In the right figure we plot $\a_{ST}$ as a function of $L^{-2}$ for $14\le L\le 34$. The shown data fits well to the functional form $\a_{ST}(L)=0.80417-0.37508L^{-2}$, suggesting that $\a_c \approx 0.804$.
}
\label{fig:crossing}
\end{figure*}

To find what happens when $J^\prime=0$, we resort to a numerical exact diagonalization approach and find the ground state as we vary $J^\prime/J$ from $1/2$ towards $0$. For these calculations, we set $J=1$ as we vary $J^\prime$. The behavior of the lowest three energy eigenvalues $E_0$, $E_1$, and $E_2$ for lattice sizes up to $L=20$ are shown in \cref{fig: NNN GMM gap}. When $J^\prime/J = 1/2$ we expect $E_1-E_0 = 0$ due to the exact degeneracy of the ground state and $E_2-E_0 \neq 0$, for all lattice sizes. However, this is not guaranteed for smaller 
values of $J^\prime$. Indeed as the figure shows, the degeneracy of the ground state is lifted for smaller values of $J^\prime$ on small lattices, but the gap $E_1-E_0$ seems to close as the lattice size increases, while $E_2-E_0$ remains non-zero even at $J^\prime = 0$. This provides strong evidence that the entire region between $J^\prime/J=0$ and $J^\prime/J=1/2$ has a dimerized two-fold degenerate ground state. Based on this result, we conclude that the TFGMM as given in \cref{eq:TFGMM-1} is VBS ordered at $\alpha=0$.

This finding of the VBS order at $\a = 0$ is consistent with the phase transition we found in the $\alpha \lesssim 1$ regime, suggesting that the TFGMM contains at least two phases, a VBS phase when $\a \approx 0$ and a $k=1$ WZW phase at $\a \approx 1$ with a phase transition at some critical coupling $0 < \a_c < 1$ in the universality class of a similar transition in the frustrated next-to-nearest neighbor $J_1-J_2$ quantum spin-half chain. It is also interesting that the next-to-nearest neighbor frustration is naturally introduced in a nearest-neighbor model via an onsite transverse field.

\section{Numerical Results}
\label{sec-4}

In this section, we report large-scale numerical studies of the TFGMM using both exact diagonalization and density matrix renormalization group (DMRG) to confirm the qualitative phase diagram discussed in the previous section. In particular, we locate the value of $\a_c$. 

We first look at the low energy spectrum of TFGMM in PBC as a function of $\a$. As discussed in \cref{subsec: gapped phase}, at $\alpha=0$ we have two degenerate ground states associated with the VBS phase and the expected finite gap to excitations. It is well known from previous studies \cite{Okamoto1992,Liu:2020ygc} that the transition between the dimerized phase and the $k=1$ WZW model can be detected by studying the lowest five energy eigenvalues. In the dimerized phase, one expects two low-lying singlets and a gapped triplet. While in the $k=1$ WZW phase, the low-lying spectrum is a unique singlet and a gapped triplet. Thus the phase transition roughly occurs when the singlet on the dimerized side crosses the triplet and becomes higher in energy. The degeneracy between the singlet and the triplet at the critical point is related to the dynamical enhancement of the $\SU(2)$ symmetry to $\SU(2)\times \SU(2)$ at the critical point due to the vanishing of a marginal operator there.

In \cref{fig:crossing}, on the left side, we plot the singlet gap $E_1-E_0$ and the triplet gap $E_2-E_0$ as a function of $\a$ for a lattice size of $L=10$. We observe that as $\a$ increases, the triplet crosses the singlet around $\a_{ST} (L=10)\approx 0.801$.
%we see a crossing in the spectrum between the quasi-degenerate singlet and the first triplet excitation.  Fig.\ref{fig: DMRG PBC N=10 intersection} shows that the lowest two excitation energies at $L=10$ intersect when the value of the transverse field parameter is $\a_{ST}\approx 0.801$.
We have studied how the crossing point $\a_{ST}(L)$ changes for increasing $L$ for $14\le L\le 34$ using DMRG. These results are plotted on the right side of \cref{fig:crossing}. Since the data fits well to the form $\a_{ST}(L)=0.80417-0.37508L^{-2}$ we 
estimate the critical point to be $\a_c\approx 0.804$.
We note that the $L^{-2}$ dependence of $\a_{ST}(L)$ is well-known for $J_1$-$J_2$ spin-$\ha$ chain \cite{Okamoto1992,PhysRevLett.75.1823}, thus, given our analysis in \cref{subsec: degen pert}, the right-hand plot in \cref{fig:crossing} is expected.

Exact diagonalization on small values of $L$ with PBC reveals another interesting property of the ground state $\ket{\j_0}$ and the 1st excited singlet state $\ket{\j_1}$. They are eigenstates of the translation operator with eigenvalues $+1$ and $-1$, respectively if $L\in 4\mathbb{Z}$ and with eigenvalues $-1$ and $+1$, respectively if $L-2\in 4\mathbb{Z}$. Further, both are eigenstates of the global $\mathbb{Z}_2$ spin-flip symmetry at $\a=0$ as described in \cref{app-1}, with eigenvalue $+1$. This means that this spin-flip symmetry remains unbroken, which is unlike what was recently found \cite{Liu:2021otq}.

We have also studied the TFGMM using DMRG for various values of $L$ in the range $28\le L\le 148$ with OBC since those calculations are easier compared to PBC.
More specifically, we wrote the DMRG algorithm using the ITensor software library in Julia \cite{ITensor,ITensor-r0.3}. 
In these studies, we focused on measuring the connected correlation functions of local operators $O_j$ in the ground state.  Here $j$ represents the lattice site and we have focused on the correlation function between sites $j=L/4$ and $j=3L/4+1$ defined as 
\begin{align}
G_O(L) \ =& \ \langle \Psi_0 |O_{j=L/4} O_{j=3L/4+1} |\Psi_0 \rangle \nonumber \\
\ &- \ \langle \Psi_0 |O_{j=L/4}|\Psi_0\rangle \langle \Psi_0 |O_{j=3L/4+1} |\Psi_0\rangle,\label{eq:correlation def}
\end{align}
where $|\Psi_0\rangle$ is the ground state approximated as a matrix product state (MPS).

\begin{figure}[ht]
\centering
\includegraphics[width=0.48\textwidth]{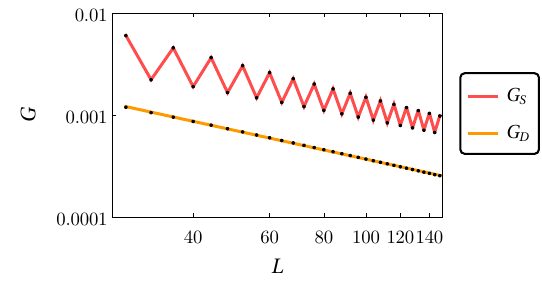}
\caption{\label{fig:new-powerlaws-1}
The plot of the correlation functions $G_S$ and $G_D$, as defined in \cref{eq:correlation}, obtained using DMRG at $\a=0.8$. The DMRG data are shown as dots, whereas the solid lines are the power-law fits obtained using a nonlinear model fit package in Mathematica \cite{Mathematica}. The fit parameters are given in \cref{tab:fitparams}. The DMRG data for $G_S$ show large oscillations which are captured by the fit function \cref{eq:correlation-d}.}
\end{figure}

\begin{figure}[ht]
\centering
\includegraphics[width=0.48\textwidth]{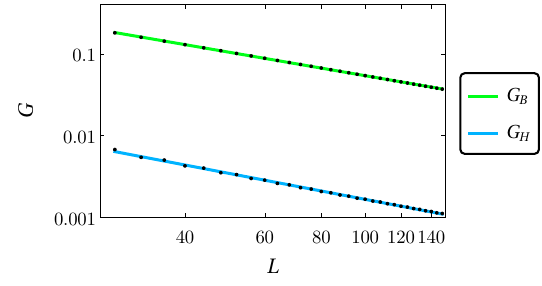}
\caption{\label{fig:new-powerlaws-2}
The plot of the correlation functions $G_B$ and $G_H$, as defined in \cref{eq:correlation}, obtained using DMRG at $\a=0.8$. The DMRG data are shown as dots whereas the solid lines are the power-law fits obtained using a nonlinear model fit package in Mathematica. The fit parameters are given in \cref{tab:fitparams}. The DMRG data for $G_H$ show small oscillations about the blue fitting curve.}
\end{figure}

\begin{table}[ht]
\centering
\renewcommand{\arraystretch}{1.4}
\setlength{\tabcolsep}{3pt}
\begin{tabular}{ cccccc}
\TopRule
$A_D$ & $P_D$ & $A_B$ & $P_B$ &   $A_H$ & $P_H$ 
\\
\MidRule
0.027(3) & 0.931(2) & 4.36(2) & 0.954(1) &  0.214(6) & 1.055(7)  \\
\TopRule
& $A_S$ & $P_S$ &  $C_S$ & $P'_S$\\
 \MidRule
& 0.1114(6) & 0.981(1)  &  -0.219(4) & 1.441(5) \\
\BotRule
\end{tabular}
\caption{\label{tab:fitparams} Parameter fit for $\alpha=0.8$. We used Mathematica's NonlinearModelFit function to fit the DMRG data to the expressions in \cref{eq:correlation}. The numbers in the parentheses are the standard error of the fitting function, they do not reflect the DMRG algorithm's error.}
\end{table}

% \begin{table}[ht]
% \centering
% \setlength{\tabcolsep}{3pt}
% \begin{tabular}{ cccccc}
% \hline
% $A_D$ & $P_D$ & $A_B$ & $P_B$ &   $A_H$ & $P_H$ 
% \\
% \hline
% 0.027(3) & 0.931(2) & 4.36(2) & 0.954(1) &  0.214(6) & 1.055(7)  \\
% \hline
% & $A_S$ & $P_S$ &  $C_S$ & $P'_S$\\
% \hline
% & 0.1114(6) & 0.981(1)  &  -0.219(4) & 1.441(5) \\
% \hline
% \end{tabular}
% \caption{\label{tab:fitparams} Parameter fit for $\alpha=0.8$. We used Mathematica's NonlinearModelFit function to fit the DMRG data to the expressions in \cref{eq:correlation}. The numbers in the parentheses are the standard error of the fitting function, they do not reflect the DMRG algorithm's error.}
% \end{table}

% \begin{figure}[b]
% \centering
% \includegraphics[width=0.48\textwidth]{Picture/cor_power_law.pdf}
% \caption{\label{fig:powerlaws}
% \RKK{Should the legends be corrected? The captions are all too short. need more descriptive details} Correlation functions discussed in the text at $\a=0.8$ which is estimated to be very close to the critical point where we expect power law decays of the form $G_O(L) \sim 1/L^{P_O}$. Using a linear model fit package in Mathematica we estimate $P_S \approx 1.0893(8)$, $P_D \approx 0.931(2)$, $P_H \approx 1.052(3)$ and $P_B \approx 0.954(1)$. \SRX{Shall we get rid of this figure?}}
% \end{figure}

The local operators we have studied include the spin operator $S^z_j$ at site $j$, dimer operator $D_j$ centered at $j$, bond energy $H_{j}$ between sites $j,j+1$, and the bond energy difference $B_j$ centered at $j$. These are defined quantitatively as
\begin{align}
    S^z_j & =(-1)^{j+1}M^-_{z,j}\,,\\
    D_j& =\frac{1}{2}(-1)^j(S^z_j S^z_{j+1}-S^z_{j-1}S^z_j)\,,\\
    H_{j}& =\sqrt{1-\a^2}\g^\m_j \g^\m_{j+1}+\frac{\a}{2}(\g^5_j +\g^5_{j+1})\,,\\
    B_j & = \ha(-1)^j(H_{j}-H_{j-1})\,,
\end{align}
where $M^-_{z,j}$ is $z$-component of the $\mathrm{SU}(2)_-$ generator at site $j$, as defined in \cref{app-1}. We will label the four correlation functions obtained using these four operators as $G_S$, $G_D$, $G_H$, and $G_B$.

For these studies, we focused on $\a=0.8$ which, according to our estimate above, is very near the critical point $\a_c$. At the critical point, we expect all correlation functions to be well-approximated by the following ansatze\footnote{Note that \cref{eq:correlation-b,eq:correlation-c} are slightly different from the form of \cref{eq:correlation def}.}:
\begin{subequations}
\begin{align}
G_D&:=\langle D_{L/4}D_{1+3L/4} \rangle-\langle D_{L/4}\rangle\langle D_{1+3L/4}\rangle \approx \frac{A_D}{L^{P_D}}\,, \label{eq:correlation-a}
\\
G_B&:=\langle B_{L/4}B_{3L/4-1}\rangle-\langle B_{L/4}\rangle\langle B_{3L/4-1}\rangle\approx \frac{A_B}{L^{P_B}}\,,\label{eq:correlation-b}\\
G_H&:=\langle H_{L/4}H_{3L/4}\rangle-\langle H_{L/4}\rangle\langle H_{3L/4}\rangle \approx \frac{A_H}{L^{P_H}}\,,\label{eq:correlation-c}\\
G_S&:= \langle S_{L/4}S_{1+3L/4} \rangle-\langle S_{L/4}\rangle\langle S_{1+3L/4}\rangle\nonumber\\
  &\approx \frac{A_S}{L^{P_S}}+(-1)^{L/4}\frac{C_S}{L^{P
  '_S}}\,.
\label{eq:correlation-d}
\end{align}
\label{eq:correlation}
\end{subequations}
From conformal field theoretic arguments we 
expect $P_S=P_D=P_H=1$ \cite{Affleck:1988px,Liu:2020ygc}. This also suggests that $P_B = 1$. In \cref{fig:new-powerlaws-1,fig:new-powerlaws-2} we plot these correlation functions at $\a=0.8$ obtained using DMRG as a function of $L$. The fit parameters obtained using a non-linear fit to the form \cref{eq:correlation} are given in \cref{tab:fitparams} and the fit functions are shown as solid lines in the figures. 
% \SCH{Check out the sentences below?} 
The powers $P_D$, $P_B$, $P_H$, and $P_S$ are close to the expected value of $1$. Moreover, $G_S$ shows oscillations qualitatively expected in \cref{eq:correlation-d} while the remaining three correlations seem to be captured by the leading terms. We believe the discrepancy between the expected results and our fit parameters is because our analysis is done at a fixed $\a=0.8$ which is not exactly at the critical point. Extracting the exact parameters at the critical point typically requires a more careful finite-size scaling analysis, which was not the goal of our work. Instead, our goal was to provide evidence of a quantum phase transition, and based on the evidence we have provided here there is little doubt that it belongs to the same universality class as the phase transition in the frustrated next-to-nearest neighbor spin-half anti-ferromagnetic chain.

\section{Conclusions}
\label{sec-5}

In this work, we have introduced a new class of quantum spin models with $\SO(4)$ symmetry realized using a projective $\Spin(4)$ Hilbert space. We hope that these models, which we refer to as $\gamma$-matrix models (or GMMs), potentially have a rich phase structure with exotic second-order critical points where quantum field theories with topological terms can emerge. To demonstrate this possibility, in this work, we used DMRG to study one of the simplest GMMs in one spatial dimension. Interestingly, even though our model only has nearest-neighbor interactions, we argued that it naturally maps to the well-known next-to-nearest-neighbor, frustrated quantum spin-half anti-ferromagnetic chain. We provided analytic and numerical evidence that our model has a quantum critical point between the critical $k=1$ WZW phase and the dimerized phase, analogous to the frustrated quantum spin-half anti-ferromagnetic chain. 

In the future, we plan to study other GMMs in higher spatial dimensions. While such models can naturally be studied on quantum computers and quantum simulators, some of them are free of sign problems and can be studied using efficient worm-type algorithms. These models will naturally have an $\SO(4)$ symmetry realized projectively on Hilbert spaces containing $\Spin(4)$ representations. It seems likely to us that exotic critical points can arise which are naturally described by quantum field theories with new types of topological terms. To discover them, a systematic study of several models in this space may be needed.

%The authors acknowledge support from NSF DMR-2312742 (RKK).

\section*{Acknowledgments}

We thank T. Bhattacharya, H. Liu, H. Katsura, T. Senthil, H. Singh, and C. Wang for helpful discussions. SC and RXS are supported in part by the U.S. Department of Energy, Office of Science, Nuclear Physics program under Award No. DE-FG02-05ER41368. SC and RXS were also partially supported by the U.S. Department of Energy, Office of Science---High Energy Physics Contract KA2401032 (Triad National Security, LLC Contract Grant No. 89233218CNA000001) to Los Alamos National Laboratory via a Duke subcontract when some of this work was performed. RKK was supported by the NSF under Award No. DMR-2312742. This research was supported in part by grant NSF PHY-2309135 to the Kavli Institute for Theoretical Physics (KITP) (RKK, SCH).

\bibliography{ref.bib}

\appendix

\section{Gamma matrices and the \texorpdfstring{$\SO(5)$}{Lg} group}
\label{app-1}

The five $4\times 4$ Dirac matrices $\g^\m,\m=1,2,3,4,5$ satisfy the Clifford algebra defined by relation
\begin{align}
    \{\g^\m,\g^\n\}=2\d^{\m\n}\unitmat\,,
\end{align}
where $\unitmat$ is the $4\times 4$ unit matrix. While there are several choices for the matrix representation of $\g^\m$, we will choose the Weyl basis given by
\begin{align}
    \g^j & =-\s^2\otimes\s^j=\begin{pmatrix} 0 & i\s^j\\
    -i\s^j & 0
    \end{pmatrix}\,,~~ j=1,2,3 \\
    \g^4 & =\s^1\otimes\s^0=\begin{pmatrix} 0 & \s^0\\
   \s^0 & 0
    \end{pmatrix}\, ,\\
    \g^5 & =-\g^1\g^2\g^3\g^4=\s^3\otimes\s^0=\begin{pmatrix} \s^0 & 0\\
    0 & -\s^0
    \end{pmatrix}\,,
\end{align}
where $\s^j$ are the usual Pauli matrices, $\s^0$ is the $2\times 2$ unit matrix and the $0$ stands for a $2\times 2$ zero matrix. Note that these gamma matrices are Hermitian: $(\g^\m)\dg=\g^\m$.

It is well known that the four-dimensional Hilbert space on which the $\g^\m$'s act can be viewed as a representation of the $\SO(5)$ group. Indeed the ten Hermitian matrices
\begin{align}
    \s^{\m\n}:=\frac{i}{4}[\g^\m,\g^\n]\, ,
\label{eq:gmndef}
\end{align}
that satisfy the $\mathfrak{so}(5)$ Lie algebra
\begin{align}
[\s^{\m\n},\s^{\k\l}]\ =\ i (\d^{\n\k}\s^{\m\l} - \d^{\n\l}\s^{\m\k} - \d^{\m\k}\s^{\n\l} + \d^{\m\l}\s^{\n\k}).
\label{eq:spin5alg}
\end{align}
This is the spinorial irreducible representation (irrep) of $\SO(5)$ or equivalently $\Spin(5)$ which is the simply-connected double-cover of $\SO(5)$. Thus, we can also view the Lie algebra as $\mathfrak{so}(5)\simeq\mathfrak{spin}(5)$. The four-dimensional Hilbert space on which the Dirac matrices act can be used to construct lattice models that have various types of symmetries related to the $\SO(5)$ group and its subgroups. Here we will be focusing on the $\SO(4)$ subgroup.

Note that if we restrict $\m=1,2,3,4$ in \cref{eq:spin5alg} we obtain the $\mathfrak{so}(4)\simeq\mathfrak{spin}(4)$ Lie algebra. From this perspective, we can also view the four-dimensional Hilbert space of the Dirac matrices as representations of the $\mathfrak{spin}(4)$. But recall the Lie algebra isomorphism $\mathfrak{so}(4)\cong \mathfrak{su}(2)\oplus\mathfrak{su}(2)$. Indeed by defining
\begin{align}
   M^{\pm}_i:=-\frac{1}{2}\left(\frac{1}{2}\e^{ijk}\s^{jk}\pm \s^{i4}\right)\, ,~~~(i,j,k=1,2,3) 
\label{eq:Mdef}
\end{align}
where we assume the repeated indices are summed over, we see that the $\mathfrak{so}(4)$ algebra splits into two mutually-commuting copies of $\mathfrak{su}(2)$:
\begin{align}
    [M^{+}_i,M^{-}_j]=0\, ,~~[M^{\pm}_i,M^{\pm}_j]=i\e_{ijk}M^{\pm}_k\, .
\end{align}
For this reason, any irreducible representation (irrep) of the Lie algebra $\mathfrak{spin}(4)=\mathfrak{su}(2)\oplus \mathfrak{su}(2)$ and can be labeled by a pair of non-negative half-integers, $(s_+,s_-)$, where $s_{\pm}$ specifies the $\mathrm{SU}(2)$ irreps for the operators $\mathbf{M}^{\pm}$. If we substitute \cref{eq:gmndef} into \cref{eq:Mdef} we obtain
\begin{align}
    M^{+}_{i}=\begin{pmatrix}
        \ha\s^i & 0\\
        0 & 0    \end{pmatrix}\,,~~M^{-}_{i}=\begin{pmatrix}
        0 & 0\\
        0 &\ha \s^i
    \end{pmatrix}\,.\label{eq: M+-}
\end{align}
which shows that the four-dimensional Hilbert space on which $\g^\m$ acts is in fact a reducible representation of $\mathrm{Spin}(4)$ and splits into $(1/2,0)\oplus (0,1/2)$.

From the perspective of global properties of groups $\mathrm{Spin}(4)=\mathrm{SU}(2)\times\mathrm{SU}(2)$ is the double cover of $\SO(4)$. Interestingly, one notices that
\begin{align}
M^{\pm}_i\ =\ (\g^4)^\dagger M^{\mp}_i \g^4,
\end{align}
which means that $\{1,\g^4\}$ is a $\mathbb{Z}_2$ group that flips the two $\mathrm{SU}(2)$ subgroups in $\mathrm{Spin}(4)$ into each other. We refer to this $\mathbb{Z}_2$ as the spin-flip symmetry. Thus, along with this $\mathbb{Z}_2$ group, the $\mathrm{Spin}(4)$ group enhances to the semi-direct product group $\Pin(4) = \mathrm{Spin}(4) \rtimes \mathbb{Z}_2$. This is analogous to $\mathrm{O}(4)= \SO(4) \rtimes \mathbb{Z}_2$. Just as the spin group $\Spin(4)$ is the double cover of the special orthogonal group $\SO(4)$, the pin group $\Pin(4)$ is the double cover of the orthogonal group $\mathrm{O}(4)$.

\section{Gamma Matrix Models}
\label{app-2}

In quantum spin-half models, the two-dimensional Hilbert space on which the Pauli matrices act forms the spinorial representation of the $\SO(3)$ group. As we discussed in \cref{app-1}, the four-dimensional Hilbert space on which the Dirac matrices $\g^\m$, $\m=1,2,3,4,5$ act forms the spinorial representation of the $\SO(5)$ group. Thus, in analogy with quantum spin models constructed with the three Pauli matrices on each lattice site, we can envision a whole class of quantum models with more interesting symmetries constructed with the five anti-commuting Dirac matrices. In this appendix, we discuss some simple models constructed using the Dirac matrices and refer to them as gamma matrix models.

Since the five Dirac matrices, $\g^\m,\m=1,...,5$, transform as a $5$-vector under the $\SO(5)$ group, models of the form
\begin{align}
H \ = \ \sum_{i \neq j}\sum_{\m = 1}^5\ J^{(5)}_{ij}\g^\m_i\ \g^\m_j
\label{eq:gmm-5}
\end{align}
are invariant under $\Spin(5)$. This model is a natural extension of Heisenberg spin-half models that are invariant under $\Spin(3)\simeq \SU(2)$ symmetry. An interesting question is whether these class of models in three spatial dimensions can naturally lead to topological $\theta$-terms involving the $\SO(5)$ symmetry, in analogy to Heisenberg spin-half chains which lead to such terms involving the $\SO(3)$ symmetry.

From the perspective of $\Pin(4)$ symmetry, the four matrices $\g^\m, \m=1,2,3,4$ transform irreducibly as $4$-vectors under the $\mathrm{O}(4)$ group. Further, $\g^5$ is invariant under $\Spin(4)$ but not under $\Pin(4)$ since it breaks the $\mathbb{Z}_2$ symmetry. Thus, the model 
\begin{align}
H \ = \sum_{i \neq j} \sum_{\m=1}^4\  J^{(4)}_{ij} \g^\m_i\g^\m_j \ +\ h \sum_j \g^5_j
\label{eq:gmm-4}
\end{align}
is invariant under $\Spin(4)$ when $h\neq 0$ and not $\Pin(4)$. In this work, we study the nearest neighbor version of this model in one spatial dimension with $J^{(4)}_{ij} = J$ and argue that the model contains the well-known transition between the two phases predicted by the LSM theorem as we vary $h$. This is because breaking the $\Pin(4)$ symmetry makes the model similar to the $\SU(2)$ spin model but with induced frustrating interactions that allow for a phase transition. 

By restricting the values of $\m$ that enter in the defining Hamiltonians, like in \cref{eq:gmm-5} and \cref{eq:gmm-4}, we can get more models, many of which have local symmetries. This implies that the Hilbert space of the models decomposes into sectors that do not mix. Breaking these local symmetries by small amounts can lead to interesting physics. This feature of gamma matrix models is illustrated by considering the
\begin{align}
H \ = \sum_{i \neq j} \sum_{\m=1}^3\  J^{(3)}_{ij} \g^\m_i\g^\m_j \ +\ h \sum_j \g^5_j.
\label{eq:gmm-3}
\end{align}
Note that when $h=0$, the Hamiltonian $H$ defined in \cref{eq:gmm-3} has a local symmetry $\mathrm{U}(1)$ generated by $Q_k=i\g^4_k\g^5_k$ at every spatial site $k$, i.e., $[H,Q_k] = 0$. The set of eigenvalues of $Q_k$, which we will define as $\{b_k\}$, can be used to label Hilbert space sectors of the model that do not mix. Interestingly, in each of these sectors the Hamiltonian describes a spin-half model since $H$ also has a global $\mathrm{SU}(2)$ symmetry generated by $\s^{23},\s^{31},\s^{21}$. If we now switch on the $h$-term, the local symmetry is broken and the various local symmetry sectors can begin to mix. By choosing $J^{(3)}_{ij}$ carefully, we believe that models of this type can induce interesting long-distance physics.

\section{Exact Ground State}
\label{app-3}

Following the ideas of Majumdar and Ghosh \cite{Majumdar:1969zmb}, it is possible to find the exact ground state of a modified GMM model with a next-to-nearest-neighbor interaction, with the Hamiltonian
\begin{align}
\HNN = \sum_{j} \sum_{\m=1}^4\left(J\g^\m_j\g^\m_{j+1}+J'\g^\m_j\g^\m_{j+2}\right)\,.
\end{align}
When $J=2J'\ge 0$, we obtain the Majumdar-Ghosh gamma matrix model 
\begin{align}
\HMG =\frac{J}{4} \sum_{j} \left[ \sum_{\m=1}^4(\g^\m_{j}+\g^\m_{j+1}+\g^\m_{j+2})^2-12\right]\,.
\end{align}
%where $K_j$ is given by
%a non-negative Hermitian operator
%\begin{align}
%K_j=\sum_{\m=1}^{4}(\g^\m_{j}+\g^\m_{j+1}+\g^\m_{j+2})^2-4\,.
%\end{align}
We can show rigorously that the ground state of $\HMG$ on a periodic lattice with an even number of lattice sites $L \geq 6$ is doubly degenerate. The two linearly independent but non-orthogonal ground states are given by
\begin{align}
|\psi\rangle_\pm \ =\ \bigotimes_{i=1,3,5...} \ket{i;i\pm 1}
\label{eq:MGGS}
\end{align}
where $|i;j\rangle$ is defined as the $\Spin(4)$ singlet ground state of the two-site Hamiltonian 
\begin{align}
h_{ij} = \sum_{\mu=1}^4 \g^\m_i\g^\m_j.
\end{align}
In the Weyl basis introduced in \cref{app-1}, this is
\begin{align}
    \ket{i;j}=\ha (e^1_i\otimes e^2_j-e^2_i\otimes e^1_j-e^3_i\otimes e^4_j+e^4_i\otimes e^3_j)\,, 
\end{align}
where $e^\m_i$ is the unit vector pointing in the $\m$-th direction in the site-$i$ local Hilbert space. 

While it is easy to see that the two states given by \cref{eq:MGGS} are ground states of $\HMG$, it is subtle to prove they are the only ground states. From numerical studies, we observe that there are only two ground states for even $L$ in the range  $6\leq L\leq 20$, see Fig.~\ref{fig: NNN GMM gap}. Here we sketch steps of a rigorous proof that the ground state is two-fold degenerate for all even system sizes $L\geq 6$. 

\begin{figure}[h]
\includegraphics[width=0.48\textwidth]{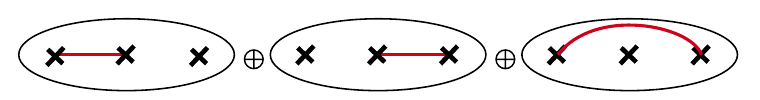}
\caption{\label{fig:G3} The $12$-dimensional ground state subspace of $K_i$ where the three sites are $i$, $i+1$, and $i+2$ shown as crosses. The red bond between the sites represents the 
$\SO(4)$ singlet state $\ket{i;j}$. An isolated cross, not connected to its neighbors, represents a four-dimensional Hilbert space of a single site. We refer to such sites as ``dangling" sites in our discussion. Thus each diagram shown represents a $4$-dimensional subspace of the $64$-dimensional three-site Hilbert space. All the vectors shown can be shown to be linearly independent.}
\end{figure}

The essential idea is to begin with the ground state subspace of the three site Hamiltonian $K_1$ where we define
\begin{align}
K_i=\sum_{\m=1}^{4}(\g^\m_i+\g^\m_{i+1}+\g^\m_{i+2})^2
\end{align}
as the Hamiltonian for the three-site problem involving the sites $i$, $i+1$, and $i+2$. We will not impose periodic boundary conditions until the end. While the full Hilbert space of the three-site problem is $64$-dimensional, one can show that the ground-state subspace is only $12$-dimensional. While an orthonormal basis of this subspace contains states that entangle all three sites, one can choose a linearly independent (non-orthogonal) basis that is rather simple to visualize. This basis can be represented pictorially using two-site singlets, as shown in \cref{fig:G3}.

\begin{figure}[h]
    \centering
    \includegraphics[width=0.48\textwidth]{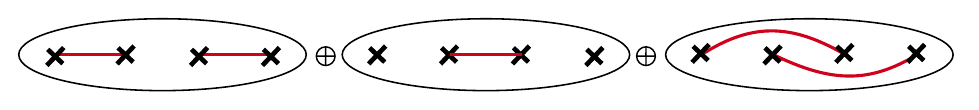}
    \caption{The figure shows the $18$ linearly independent states that span the ground state of the four-site Hamiltonian $K_1+K_2$. The bonds and the free sites have the same meaning as \cref{fig:G3}.}
    \label{fig:G4}
\end{figure}

We then add one additional neighboring lattice site and consider the four-site problem with a $256$-dimensional Hilbert space. When we consider the ground state subspace of the Hamiltonian $K_1+K_2$, the Hilbert space is reduced to $18$-dimensional subspace spanned by the linearly independent states shown in \cref{fig:G4}. One can view this reduced subspace as the intersection of the ground state subspaces $K_1$ and $K_2$ in the four-site Hilbert space, each of which is $48$-dimensional.

\begin{figure}[h]
    \centering
    \includegraphics[width=0.48\textwidth]{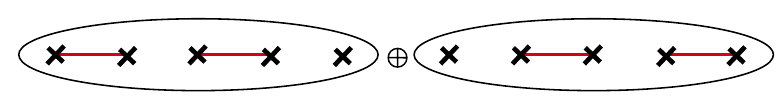}
    \caption{The ground state subspace of the five-site Hamiltonian $K_1+K_2+K_3$ is $8$-dimensional and is shown here pictorially.  \label{fig:G5}}
\end{figure}

\begin{figure}[h]
    \centering
    \includegraphics[width=0.48\textwidth]{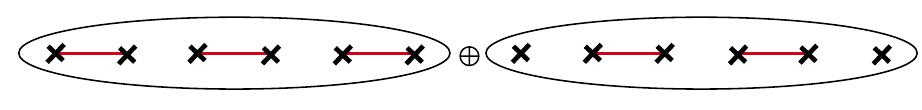}
    \caption{The ground state subspace of the six-site Hamiltonian $K_1+K_2+K_3+K_4$ is $17$-dimensional and is shown here pictorially. Note that it is no longer $18$-dimensional since one of the states shown in \cref{fig:G4} is not allowed.
    \label{fig:G6}}
\end{figure}

We can then repeat this process by adding one additional site. In the five-site problem, the intersection of the $72$-dimensional ground state subspace of $K_1+K_2$ and the $192$-dimensional ground state subspace of $K_3$ turns out to be $8$-dimensional, which is pictorially shown in \cref{fig:G5}. Going to six sites, the intersection of the $32$-dimensional ground state subspace of $K_1+K_2+K_3$ and $768$-dimensional ground state subspace of $K_4$ turns out to be $17$-dimensional, shown pictorially in \cref{fig:G6}.

Beyond six sites, as we add more sites, the analysis can be shown to repeat. In general, for $L$ sites we can show that the low-energy subspace of the Hamiltonian $K_1+K_2+...+K_{L-2}$ is $8$-dimensional if $L$ is odd and $17$-dimensional if $L$ is even. Pictorially the linearly independent states can be understood as extensions of the figures \cref{fig:G5} and \cref{fig:G6}. When $L$ is odd, there is a dangling site on the right or a dangling site on the left, with $\Spin(4)$ singlets connecting all other nearest neighbors. Each dangling site counts as four states. When $L$ is even, either the $\Spin(4)$ singlets connect all neighboring sites starting from one end, or there are two dangling sites at the ends while the remaining sites are connected as $\Spin(4)$ singlets. The former counts as one state while the latter counts as $16$ states.

If we impose periodic boundary conditions and only consider even $L\ge 6$, the ground state subspace reduces to being two-dimensional. Now there are no dangling spins allowed and the two dangling spins at the two ends are connected as a $\Spin(4)$ singlet. These are the two states given in \cref{eq:MGGS}.

\section{Lieb-Shultz-Mattis theorem}
\label{app-4}
In this section, we prove that the Lieb-Shultz-Mattis (LSM) theorem applies to our model, in that the TFGMM cannot have a unique gapped ground state in the thermodynamic limit.

The LSM theorem has two parts. First, we need to show there is a state $\ket{\j_{twist}}$ orthogonal to the unique ground state $\ket{\j_0}$ of $H_L$ for any $L$; then, we show this state's energy expectation value converges to the ground state energy $E_0$ as $L\rightarrow \infty$.
The first part will follow from standard arguments once we find a translation-invariant global $\mathrm{U}(1)$ symmetry generator whose local density operator has half-odd-integer eigenvalues. This part is independent of the specific form of the Hamiltonian. The second part deviates from the original proof of the LSM theorem \cite{LSM1961} because it depends on the specifics of the Hamiltonian. In particular, unlike the Heisenberg spin chain, the TFGMM bond operator $\hat{h}_{i,j}$ is not a product of the symmetry generators.
Nevertheless, the spectral gap can be shown to be bounded above by $\frac{1}{L}$ using a trick we learned from \cite{tasaki2020physics,koma2000spectral}.

The lattice is modeled by $\L=\mathbb{Z}_{>0}$, i.e. the set of all positive integers, and we define the length-$L$ intervals $\L_L=\{x\in \L~|~1\le  x\le L\}$ for any even integer $L$.
For each $L$, the periodic Hamiltonian is
\begin{align}
    H_L=\hat{h}_{L,1}+\sum_{j=1}^{L-1}\hat{h}_{j,j+1}\,,
\end{align}
where the symmetrized bond energy operator is
\begin{align}
   \hat{h}_{i,j}= J\g^\m_i\g^\m_{j}+\frac{h}{2} (\g^5_i+\g^5_{j})\,.
\end{align}
The ground state and ground state energy of $H_{L_i}$ are denoted by $\ket{\j_0}$ and $E_0$, respectively, where we do not explicitly show the dependence on lattice size $L$.

If there is an infinite strictly-increasing sequence of lattice sizes $L$ for which the TFGMM has a degenerate ground state, we would have obtained the desired result, and nothing else needs to be done. Thus, from now on we assume we are given a sequence of sublattices $\{\L_{L_i}|i\in \mathbb{N}\}$ with the associated periodic Hamiltonian $H_{L_i}$ such that $L_i\in 2\mathbb{Z}$, $L_i<L_{i+1}$, and the ground state $\ket{\j_0}$ of $H_{L_i}$ is unique for each sufficiently large finite $L_i$.

Define the twist operator to be
\begin{align}
    \hat{U}_{twist}=\exp\left[-i\sum_{j=1}^{L}\frac{2\p  j}{L}\hat{S}^{(z)}_j\right]\,.
\end{align}
Here
\begin{align}
    \hat{S}^{(z)}_j=3 M_{3,j}^{+}+M_{3,j}^{-}\,,
\end{align}
where $M^{\pm}_{3,j}$ are site-$j$ operators as in \cref{app-1}:
\begin{align}
    M^{\pm}_{3,j}=-\ha (\s^{12}_j\pm \s^{34}_j)\,.
\end{align}
The specific linear combinations of $M^{\pm}_{3,j}$ in the definition of $\hat{S}^{(z)}_j$ is chosen so that, in the Weyl basis, the matrix of $\hat{S}^{(z)}_j$ is diagonal with half-odd-integer eigenvalues $\pm \frac{3}{2},\pm \frac{1}{2}$.
Since $\hat{S}^{(z)}_j$ is a linear combination of $M^{\pm}_{3,j}$, the operator $\hat{S}^{(z)}_{tot}=\sum_{j=1}^{L}\hat{S}^{(z)}_j$ is a global U(1) symmetry generator of TFGMM. 
These two facts, that $\hat{S}^{(z)}_j$ has half-odd-integer eigenvalues and $\hat{S}^{(z)}_{tot}$ being a global U(1) symmetry generator of our theory, are essential to proving the first part of the LSM theorem. The rest of the argument is standard.

Define the twisted state $\ket{\j_{twist}}=\hat{U}_{twist}\ket{\j_0}$. 
If $\ket{\j_0}$ is normalized, then so is $\ket{\j_{twist}}$, because $\hat{U}_{twist}$ is unitary.
Define $E_{twist}$ as the energy expectation value with respect to $\ket{\j_{twist}}$:
\begin{align}
    E_{twist}=\langle\j_{twist} |H_L|\j_{twist} \rangle\,.
\end{align}

\begin{Lemma}\label{lemma: LSM orthogonality}
The twisted state $\ket{\j_{twist}}$ is orthogonal to the unique ground state $\ket{\j_{0}}$:
\begin{align}
    \langle \j_{twist}|\j_0\rangle=0\,. \label{eq: LSM orthogonality}
\end{align}
\end{Lemma}
\begin{proof}
    \item The translation $\hat{T}$ acts on $=\hat{S}^{(z)}_{i}$ by
    \begin{align}
        \hat{T}\dg \hat{S}^{(z)}_{j}\hat{T}=\hat{S}^{(z)}_{j-1}\,,~~\hat{T}\dg \hat{S}^{(z)}_{1}\hat{T}=\hat{S}^{(z)}_{L}
    \end{align}
    for any $j=2,3,\cdots,L$.
    It follows that
    \begin{align}
        \hat{T}\dg \hat{U}_{twist}\hat{T}&=e^{i 2\p \hat{S}^{(z)}_{L}}\hat{U}_{twist}e^{-i(2\p/L)\hat{S}^{(z)}_{tot}}\\
       &=-\hat{U}_{twist}e^{-i(2\p/L)\hat{S}^{(z)}_{tot}} \,,
    \end{align}
    where the second line follows from the fact that the eigenvalues of $2\hat{S}^{(z)}_L$ are odd-integers.
    The unique ground state must be a singlet under all symmetry operators: for the U(1) generator, $\hat{S}^{(z)}_{tot}\ket{\j_{0}}=0$, whereas it can at most gain a phase under translation, $\hat{T}\ket{\j_{0}}=e^{-i\q}\ket{\j_{0}}$. Then
    \begin{align}
        \langle \j_{0}|\j_{twist}\rangle&=\langle \j_{0}| \hat{U}_{twist}|\j_{0}\rangle\\
        &=\langle \j_{0}| \hat{T}\dg \hat{U}_{twist}\hat{T}|\j_{0}\rangle\\
        &=-\langle \j_{0}| \hat{U}_{twist}e^{-i(2\p/L)\hat{S}^{(z)}_{tot}}|\j_{0}\rangle\\
        &=-\langle \j_{0}| \hat{U}_{twist}|\j_{0}\rangle\\
        &=-\langle \j_{0}|\j_{twist}\rangle\,.
    \end{align}
    Therefore $\langle \j_0|\j_{twist}\rangle=0$.
\end{proof}
Now we will prove the second part of the LSM theorem, using the specific form of the TFGMM Hamiltonian.
We will make use of the concept of an operator norm: the operator norm $\|\hat{A}\|$ of a self-adjoint operator $\hat{A}$ is the smallest upper bound to the magnitude of the operator's eigenvalues.

\begin{Lemma}\label{lemma: LSM}
For all $L\ge 10$,
\begin{align}
    0\le E_{twist}-E_0\le \frac{C}{L}\label{eq: LSM lemma}
\end{align}
where $C=40\p^2|J|$.
\end{Lemma}
\begin{proof}
The upper bound of the spectral gap $E_{twist}-E_0$ is typically derived by evaluating the operator norm of some local operator $\hat{A}_j$ constructed out of $\hat{h}_{i,i+1}$ and $\hat{U}_{twist}$.
% However, sometimes the choice of these local operators is not too obvious. E.g., consider the transverse-field $\mathrm{O}(3)$ Heisenberg model with nearest-neighbor interaction:
% \begin{align}
%     \hat{h}_{i,j}=J\sum_{k=1}^{3}\hat{S}^{(k)}_{i}\hat{S}^{(k)}_{j}+\frac{h}{2}(\hat{S}^{(3)}_{i}+\hat{S}^{(3)}_{j})\,.
% \end{align}
% The easiest way to derive an upper bound of the spectral gap is as follows:
% \begin{align}
% E_{twist}-E_0& =\langle \j_0|\hat{U}\dg_{twist} [H_L,\hat{U}_{twist}]|\j_0\rangle\\
% & =\sum_{j}\langle \j_0|\hat{U}\dg_{twist} [\hat{h}_{j,j+1},\hat{U}_{twist}]|\j_0\rangle\\
% & \le \sum_{j}\|\hat{U}\dg_{twist} [\hat{h}_{j,j+1},\hat{U}_{twist}]\|\\
% & = \sum_{j}|J| C \sin(\p/L)\\
% & \le  \sum_{j}|J| C \frac{\p}{L}\\
% & = |J|\p C\,,
% \end{align}
% where $C>0$ is some constant depending on the spin $S$ of the Heisenberg chain, and the fifth line holds for any $L>\p$. This is an upper bound for the spectral gap, but it is not strong enough to prove the second part of LSM theorem. 
We note that
\begin{align}
    0\le E'_{twist}-E_0& =\langle \j_0|\hat{U}_{twist} [H_L,\hat{U}\dg_{twist}]|\j_0\rangle\,,
\end{align}
where 
\begin{align}
    E'_{twist}=\langle \j'_{twist}|H_L|\j'_{twist}\rangle\,,\\
    \ket{\j'_{twist}}=\hat{U}\dg_{twist}\ket{\j_0}\,.
\end{align}
Hence, %\SRX{The 2nd line shouldn't be changed, contrary to what we discussed earlier. Should not put in a factor of half, it would be incorrect. One can check in Mathematica that the difference of the two operators is non-zero.}
\begin{align}
    E_{twist}-E_0&=\langle \j_0|\hat{U}\dg_{twist} [H_L,\hat{U}_{twist}]|\j_0\rangle\\
    & \le \langle \j_0|\hat{U}\dg_{twist} [H_L,\hat{U}_{twist}]|\j_0\rangle\nonumber\\
    &~~~~+\langle \j_0|\hat{U}_{twist}[H_L,\hat{U}\dg_{twist}]|\j_0\rangle\\
    &=\langle\j_0|  [\hat{U}\dg_{twist}[H_L,\hat{U}_{twist}]]|\j_0\rangle\\
    &=\sum_{j=1}^{L} \langle\j_0|  [\hat{U}\dg_{twist}[\hat{h}_{j,j+1},\hat{U}_{twist}]]|\j_0\rangle\nonumber\\
    &\le \sum_{j=1}^{L} \| \hat{A}_j\|\,,
\end{align}
where, $\hat{A}_j= [\hat{U}\dg_{twist}[\hat{h}_{j,j+1},\hat{U}_{twist}]]$ with $\hat h_{L,L+1}\equiv \hat h_{L,1}$.
We now show that, for all sufficiently large $L$,
\begin{align}
    \|\hat{A}_j\|=\|[\hat{U}\dg_{twist},[\hat{h}_{j,j+1},\hat{U}_{twist}]]\| \le \frac{C}{L^2}
\end{align}
for some constant $C>0$ that doesn't depend on $j,L$.
We find that the non-zero eigenvalues of $\hat{A}_j$ are 
\begin{align}
   \pm 8 J \sin^2\left(\frac{\p}{L}\right)\,,~\pm 8 J \sin^2\left(\frac{2\p}{L}\right)\,,\\
   \pm 8J\left[3+2\cos\left(\frac{2\p}{L}\right)\right]\sin^2\left(\frac{\p}{L}\right)\,,\\
   \pm 8J\sin\left(\frac{3\p}{L}\right)\sin\left(\frac{\p}{L}\right)\,.
\end{align}
These eigenvalues are bounded above by $\frac{40\p^2|J|}{L^2}$ for any $L\ge 3\p$, because 
\begin{align}
    0\le \sin(x)\le x\,, ~~~\forall x\in [0,1]\,.
\end{align}
Therefore, 
\begin{align}
    0\le E_{twist}-E_0\le\sum_{j=1}^{L}\frac{40\p^2|J|}{L^2}=\frac{40\p^2|J|}{L}
\end{align}
for all $L\ge 10$.
\end{proof}

\cref{lemma: LSM orthogonality} and \cref{lemma: LSM} lead to the LSM theorem:
\begin{Th}
    The TFGMM cannot have a unique gapped ground state in the thermodynamic limit.
\end{Th}
\begin{proof}
    Assuming the TFGMM has a unique ground state in the thermodynamic limit, we will show that it cannot be gapped. If the unique ground state were gapped in the thermodynamic limit, then all states orthogonal to $\ket{\j_0}$ must have energy expectation higher than $E_0$ by at least $\D E>0$. However, by \cref{lemma: LSM orthogonality}, there is a normalized state $\ket{\j_{twist}}$ orthogonal to $\ket{\j_0}$ for any $L$, and, by \cref{lemma: LSM}, its energy expectation $E_{twist}$ converges to $E_0$ as $L\rightarrow\infty$. Therefore, the TFGMM cannot have a unique gapped ground state in the thermodynamic limit.
\end{proof}

While we proved the second part of the theorem specifically for the TFGMM, a more general proof can be extended to a large class of $\gamma$-matrix models without making assumptions of the specific form of the model, see \cite{tasaki2020physics}.

\end{document}